\documentclass[prb,twocolumn,showpacs,byrevtex,floatfix,superscriptaddress]{revtex4}
\usepackage{amssymb,afterpage,amsmath}
\usepackage{graphicx}

\begin{document}

\title{\boldmath Spin-lattice interaction in the quasi one-dimensional
helimagnet LiCu$_2$O$_2$\unboldmath}
 \author{L\'aszl\'o Mih\'aly}
 \affiliation{Department of Physics and Astronomy, Stony Brook
University, Stony Brook, NY 11794-3800, USA}
 \affiliation {Electron Transport Research Group of the Hungarian Academy of Science and
Department of Physics, Budapest University of Technology and
Economics, 1111 Budapest, Hungary}
 \author{Bal\'azs D\'ora}
 \author{Andr\'as V\'anyolos}
 \affiliation{Electron Transport Research Group of the Hungarian
Academy of Science and Department of Physics, Budapest University of
Technology and Economics, 1111 Budapest, Hungary}
 \author{Helmuth Berger}
 \author{L\'aszl\'o Forr\'o}
 \affiliation{EPFL, Lausanne, CH-1015 Switzerland}
 \date{\today}
\begin{abstract}
The field dependence of the electron spin resonance in a helimagnet,
LiCu$_2$O$_2$,  was investigated for the first time. In the
paramagnetic state a broad resonance line was observed corresponding
to a $g$-factor of 2.3. In the critical regime around the
paramagnetic to helimagnetic phase transition the resonance broadens
and shifts to higher frequencies. A narrow signal is recovered at
low temperature, corresponding to a spin gap of 1.5~meV in zero
field.  A comprehensive model of the magnons is presented, using
exchange parameters from neutron scattering (T. Masuda \textit{et
al.} Phys. Rev. B \textbf{72} 014405), and the spin anisotropy
determined here. The role of the quantum fluctuations is
discussed.
\end{abstract}
\maketitle

Ground state solutions of the anti\-fer\-ro\-mag\-ne\-tic He\-i\-sen\-berg model
range from the N\'eel state  to valence bond solids and resonating
valence bonds, depending on the interactions, the dimensionality and
connectivity of the system and other factors. \cite{review, lemmens}
When the ground state has long range N\'eel order, the excitation
spectrum is gapless at $q=0$. This follows from the rotational
invariance of the Hamiltonian: the creation of long wavelength
excitations, where the relative angle of the neighboring spins
changes only by an infinitesimal amount, costs very little energy.
Gapless excitations, on the other hand, cause strong quantum
fluctuations, suppressing or destroying the N\'eel
state.\cite{anderson} The spin-orbit coupling leads to new terms in
the Hamiltonian, including the single ion anisotropy, the exchange
anisotropy, the Dzyaloshinskii-Moriya coupling and others. The
coupling between the direction of the spins and the lattice removes
the rotational invariance in many real materials, creates a gap in
the spin wave spectrum, and contributes to the stability of the
quasi-classical ground state.

In this work we investigated the magnetic excitations in a quasi
one-dimensional helimagnet,  LiCu$_2$O$_2$, in fields of 0~T-14~T
and at temperatures 2.5~K-60~K. The main result is that in the
ground state two of the three magnon branches are gapped with
$\Delta=11.8$~cm$^{-1}$=1.5~meV. Our calculations reveal that this
gap is due to a spin-lattice coupling of $D=0.079$~meV.  We show
that the proper treatment of the helical order of the spins explains
the order-of-magnitude difference between the coupling parameter and
the gap, enabling this weak coupling to  reduce the quantum
fluctuations significantly.
\begin{figure}
\includegraphics[width=5cm]{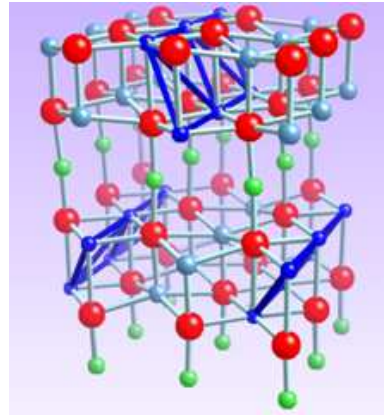}
\caption{(Color online.) Crystal structure of LiCu$_2$O$_2$, based
on Ref. \onlinecite{berger}. The magnetic Cu$^{2+}$ ions are dark
blue, the non-magnetic copper is green, the lithium is light blue
and the oxygen is red. The dark blue bonds emphasize the triangular
spin ladder.}
\end{figure}

LiCu$_2$O$_2$ crystallizes in an orthorhombic structure with the
space group Pnma and lattice constants $a=5.730$~\AA, $b=2.86$~\AA,
$c=12.417$~\AA\cite{berger}. The Cu$^{2+}$ ions carry magnetic
moments with spin 1/2, and they form quasi one-dimensional chains
with a zig-zag "ladders" along the $b$ direction (Fig. 1).  The
material has two characteristics relevant to frustrated quantum
magnets: the exchange interaction goes beyond the first neighbors,
and the structure has a triangular motif susceptible to frustration.

Quantum effects and fluctuations are expected to be important since
the spins are 1/2 and the system is quasi one-dimensional. Indeed,
in an early work Zvyagin \textit{et al.} interpreted the broadening
and disappearance of the ESR signal in terms of a "dimer liquid"
state below the 24~K phase transition \cite{zvyagin}. However,
instead of a quantum liquid state, NMR, $\mu$SR and neutron
scattering results revealed a quasi-classical helical spin order
along the chains, with a wavevector $\bf Q$ incommensurate to the
lattice \cite{gippius,roessli,masuda}.  All spins were found to be
parallel to the $a-b$ plane.  In the $\bf q=\pm \bf Q$ region the
"acoustic" magnon branches were observed, but the energy resolution
was not sufficiently high to exclude or measure a possible gap
\cite{masuda2}.  ESR has the resolution to address this question,
and the application of the external static magnetic field can be
used to explore the spin dynamics further. Note that in ESR the
magnetic excitation is at $q=0$, but in the presence of a magnetic
superlattice it couples to the modes at $\bf q=\pm \bf Q$ as well.

The measurements were performed on an oriented single crystal sample
of dimensions  3~mm$\times$6~mm$\times$0.5~mm.  According to the
Laue diffractogram, the $c$ axis is along the shortest dimension of
the sample; the $a$ and $b$ directions are parallel to the edges of
the slab. There is a twinning in the $a-b$ plane, as typical of most
LiCu$_2$O$_2$ samples.  Spin resonance was detected in the
transmission of the far infrared light, measured at Stony Brook
University's high magnetic field/infrared facility at the U12 IR
beamline of the National Synchrotron Light Source. The light
propagated parallel to the static magnetic field, and passed through
the sample along $c$ direction. The polarization of the incident
light was controlled and set to several directions within the $a-b$
plane.

Figure 2 shows the temperature dependence of the spin resonance at
12~T field.   The raw transmission curves were normalized to the
transmission of the sample in 0~T at 25~K, when no spin resonant
absorption is expected.  The oscillations seen in the baseline of
the frequency dependence are residuals of the interference fringes
seen in the raw spectra. These fringes are common and well
understood for samples of plane-parallel geometry.
\begin{figure}

\includegraphics[width=6.5cm]{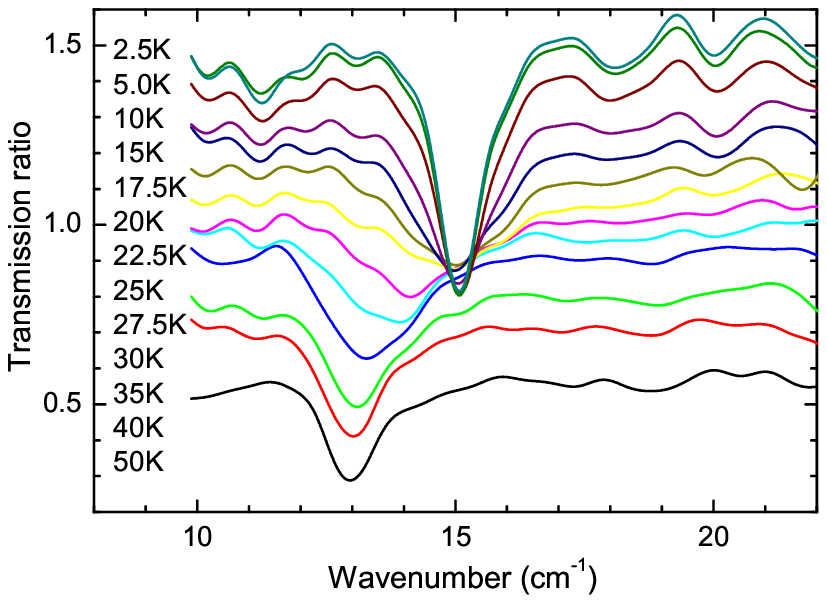}
\hspace*{-6mm}
\includegraphics[width=7.5cm]{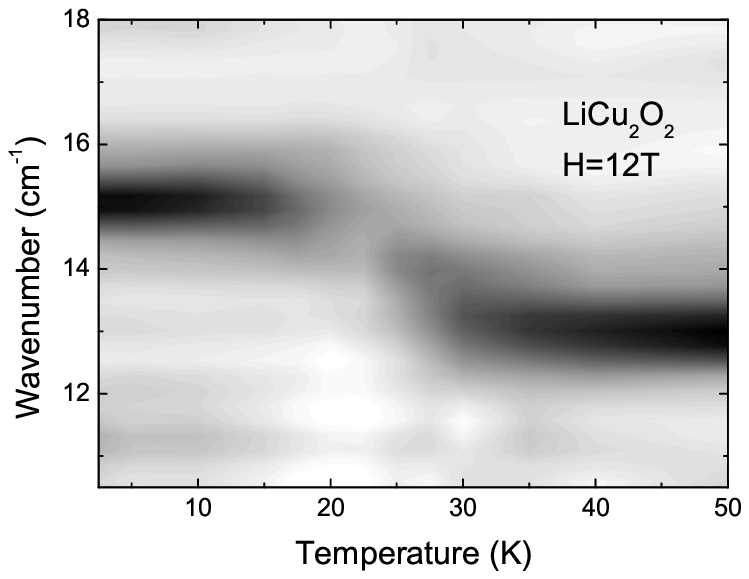}

\caption{Temperature dependence of the ESR signal at 12T magnetic
field.  The measured transmissions, relative to the transmission in
zero field, 25K, are shown.  Colors indicate the temperature. In the
lower panel the same data is assembled into a two-dimensional
intensity map.}
\end{figure}

At high temperatures a broad resonance at the "free spin" frequency
of $\hbar \omega = g \mu_B H$ with a $g$ factor of $g=2.3$ is
observed. The ESR line starts to broaden below 30~K; all of this is
in agreement with earlier ESR investigations\cite{zvyagin,
vorotinov, vorotinov2}. In the critical regime around 22~K-24~K the
ESR line is about 3~T-4~T broad (full width at half maximum), and it
shifts to higher frequency. Below 15~K a narrower signal is
recovered at a markedly higher frequency. This signal,
characteristic to the development of a spin gap in the ordered
phase, has not been seen in earlier spin resonance works
\cite{zvyagin, vorotinov, vorotinov2}, due to limitations in
accessible fields and frequencies in those studies. Measurements at
several other fields (8~T,10~T and 14~T), yielded results
qualitatively similar to the behavior presented in Fig. 2, and will
be published elsewhere.  Choi \textit{et al.} suggested a transition
from helimagnetic to N\'eel order\cite{choi} at T=9~K, but there is
no indication of  a change in the ESR signal in that range.

Zvyagin and co-workers measured ESR in the paramagnetic state at
277~GHz (9.2~cm$^{-1}$) \cite{zvyagin}. They observed the broadening
and disappearance of the ESR signal, interpreting it in terms of the
formation of a spin singlet liquid state above the ordering temperature. In fact, as our data shows, the
integrated intensity of the signal remains approximately constant as
we cross the phase transition temperature.  Vorotinov \textit{et
al.} observed an "antiferromagnetic resonance" (AFMR) signal at
$\sim$30~GHz (1~cm$^{-1}$) \cite{vorotinov, vorotinov2} in zero
field, with an unexpected and unexplained magnetic field dependence
and anisotropy.  There is no information about the intensity of this
signal relative to the ESR in the paramagnetic state. However, since
nearly all of the spectral weight of the paramagnetic resonance
remains in the high field/frequency range studied here, the AFMR
signal at 1cm~$^{-1}$ must be relatively weak, and may be related to
low energy processes that are secondary to the physics of this
material.

Figure 3 shows the field dependence of the spin resonance at low
temperatures.  The three data sets in the upper panel correspond to
the different polarization states of the incident light.  Each set
of curves were obtained by dividing the measured spectrum with a
reference spectrum recorded at a temperature above the phase
transition in zero external field.  The resonance lineshapes,
frequencies and the intensities are similar for all three
polarization states.  The most striking feature is the dramatic
increase of the resonant absorption at higher magnetic fields.

To extract the field dependence of the resonant frequency we took
the average of the three sets of data, and assembled an intensity
map (Fig. 4).  In terms of simple, two-sublattice antiferromagnets
both the intensity and the field dependence are surprising.  The
intensity of the antiferromagnetic resonance should have a very weak
field dependence \cite{talbayev} only.  More importantly, the
typical field dependence of the resonance frequency in
antiferromagnets ($\omega \propto \sqrt{H^2 + H_0^2}$) can not be
fitted to the data without residual systematic deviations.

The model used for the magnon spectrum \cite{masuda2} was adopted to
evaluate the results presented in Fig. 3. We extended it to finite
magnetic fields perpendicular to the plane of the spins, and added
the terms responsible for the spin gap.  The Hamiltonian is:
\begin{eqnarray}
{\cal{H}} = \sum_{i,j} J_1 S_{i,j}S_{i+1,j} + J_2 S_{i,j} S_{i+2,j}+
 J_4 S_{i,j} S_{i+4,j}+
 \nonumber\\
 +J_\perp S_{i,j}S_{i,j+1}-g\mu_BHS^y_{i,j}+{\cal H}',
\label{hamilton}
\end{eqnarray}
where
\begin{eqnarray}
 {\cal H}'=D_s(S^y_{i,j})^2   \text{~~~~or~~~~~}
 {\cal H}'=D_{ex}S^y_{i,j}S^y_{i+1,j}
\end{eqnarray}
The indices $i$ and $j$ run along and perpendicular to the
double chains, respectively.  $J_1$ is the coupling between along
the diagonal "rungs" of the ladder, $J_2$ and $J_4$ are the nearest
neighbor and the second neighbor  couplings along the chain, and
$J_\perp$ is the interchain coupling.  The "easy plane" is
represented either by the single ion anisotropy of $D_s>0$, or by
the exchange anisotropy of $D_{ex}<0$ (the negative $D_{ex}$ is
required because the orientation of the nearest neighbor spins has
an antiferromagnetic character). Following conventions established in
the literature\cite{nagiyama,shiba}, the reference frame for the
spin components is selected so that the magnetic field $H$ is
applied along the $y$ directions and in zero field the spins are in
the $x-z$ plane (this makes the crystallographic $c$ direction
parallel to the $y$ axis). In finite magnetic field the spin
directions are on a cone whose axis is parallel to $y$.

The ground state and the spin wave excitations of this model were
found with the methods described in Refs.
\onlinecite{nagiyama,shiba,masuda2}. Following Masuda \textit{et
al.} \cite{masuda2} we used $J_1=6.4$~meV, $J_2=-11.9$~meV,
$J_4=7.6$~meV and $J_\perp=1.8$~meV, and in zero field and with no
anisotropy we reproduced the fits to the published spin wave
spectrum. After carefully tracing the factor 2's, we believe that
the coupling constants published in the preprint version of Ref.
\onlinecite{masuda2} were correct.

In general, a helical spin arrangement possesses 3 Goldstone modes,
one at $\omega(\bf 0)$ corresponding to the free rotation of the
ordered moments within the helical plane, and two degenerate ones at
$\omega(\pm\bf Q)$ associated with the tilting of the plane of the
helix\cite{shiba,essler}. The "easy plane" anisotropy generates a
finite spin gap at $\pm \bf Q$, where ${\bf Q} =
(\pi/a,\phi/b)$ is the ordering wave vector with $\phi=5.2$rad (the angle between 
subsequent spins on the same leg of the ladder). In terms of the
Fourier transform of the exchange interaction, $J({\bf
q})=2(J_1\cos(bq_b/2)+J_2\cos(bq_b)
+J_4\cos(2bq_b)+J_\perp\cos(aq_a))$, the gap is
\begin{eqnarray}
\Delta= 2S \sqrt {J^\prime D_s} \text{~~~~~~~or~~~~~~} \Delta=
2S\sqrt {J^\prime \cos (\phi/2) D_{ex}}
\end{eqnarray}
for single ion or exchange anisotropy, respectively.  The effective
exchange coupling $J'$ is defined as
\begin{eqnarray}
J^\prime=\frac{J(2{\bf Q}) + J({\bf 0})}{4} - \frac{J({\bf Q})}{2}
\end{eqnarray}
\begin{figure}
\includegraphics[width=7cm]{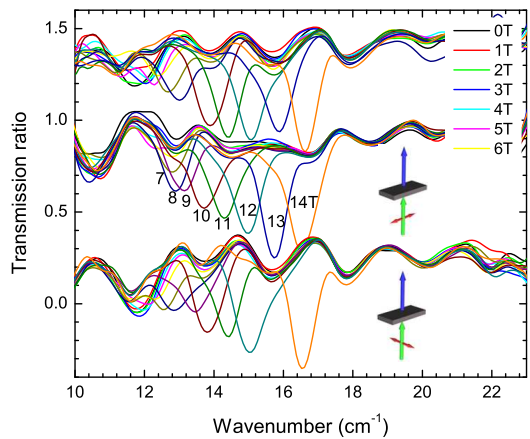}

\caption{Field dependence of the transmission at
2.5K.  The three sets of curves represent the
transmission relative to the transmission at high temperature and
zero field. Top: partially polarized light, relative to 25K. Middle:
light polarized parallel to the long axis of the sample, relative to
20K. Bottom: polarization along the short axis, relative to 25K.}
\end{figure}
\begin{figure}
\includegraphics*[width=7cm]{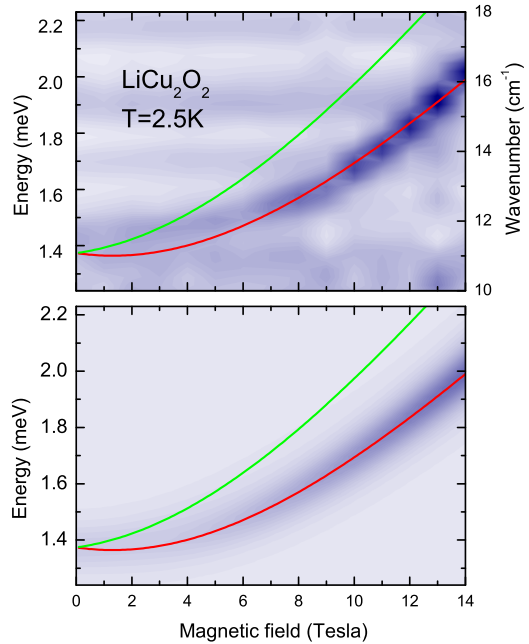}

\caption{Upper panel: Intensity map based on the average of the
three datasets shown in Fig. 3.  Lower panel: Intensity map
generated from the calculated frequencies and susceptibilities,
based on the model described in the text.  The lines represent the
resonance frequency corresponding to the two branches, also plotted
in the upper panel. }
\end{figure}
The application of external magnetic field lifts the degeneracy of
the $\pm\bf Q$ points, resulting in distinct spin gaps and different
spin susceptibilities (Fig. 4).   The continuous lines in the Figure
represent the center-frequencies of the two branches generated by
the application of the external field. The strength of the resonance
line was set according to the calculated spin susceptibility, and
lifetime effects were simulated by Lorentzian lineshapes of 0.1meV
relaxation rate, independent of field. On the lower branch, the
strong absorption at high fields gradually decreases as the field is
lowered, due to the field dependence of the susceptibility. The
susceptibility belonging to the upper branch is so small that the
the corresponding resonance signal is below the noise level.

In order to obtain the near-perfect agreement with the experiment we
adjusted two parameters:  the magnitude of the anisotropy,
$D_s=D_{ex}\cos(\phi/2)=0.079$, was set to reproduce the zero-field
gap, and the $g$-factor was $g=1.85$. Instead of using a single ion
anisotropy in Eq. \eqref{hamilton}, one may use exchange anisotropy or
Dzyaloshinskii-Moriya interaction, with very similar results
\cite{tobep}. The microscopic origin of all of these interactions is
the spin orbit coupling,  but \textit{ab-initio} calculations are
difficult. Moriya\cite{moriya} estimated the anisotropic component
of the exchange interaction in the order of $D \sim(\Delta g / g)^2
J$, where $\Delta g $ is the g-factor shift in the paramagnetic
state and $\Delta g / g$ is a measure of the spin orbit coupling. A
spin-lattice coupling in the range of $D\sim 100$~$\mu$V is
therefore consistent with the $g$-factor ranging between 1.95 and
2.3 in the paramagnetic state.\cite{vorotinov, vorotinov2} The
energy gap scales linearly with the spin orbit and exchange
couplings, $(\Delta\sim \Delta g / g) J$.

Quantum fluctuations suppress the sublattice magnetic order by the
amount of $ \Delta S = \int \frac{g(\bf q)}{\omega (\bf q)} d^D q $
(here $D$ is the dimensionality of the lattice,  $\omega(\bf q)$ is
the magnon excitation energy at wavevector $\bf q$, and $g(\bf q)$
is expressed in terms of $J(\bf q)$).\cite{anderson, shiba}.  With
no spin-lattice coupling we obtained $ \Delta S = 0.150$, a 30\%
drop relative to spin $S=1/2$, which is to be contrasted with
$\Delta S=0.197$ for a two dimensional antiferromagnet. In the
presence of the easy plane term the corresponding value is $ \Delta
S = 0.125$. The coupling to the lattice makes the quasi-classical
state more stable. Further reduction of $\Delta S$ is seen in the
presence of static magnetic field. Further evidence for reduced
sublattice magnetization follows from $g$-factor determined in the
fit.  The macroscopic energy scale of the interaction with the
magnetic field is determined by the product $g\mu_B H S_{eff}$. In
our treatment the quantum effects were neglected, and instead of a
reduced $S$ the quantum correction appears as a reduction of $g$.

In conclusion, we established that the spin wave spectrum of
LiCu$_2$O$_2$ has a gap of 1.5meV, and interpreted the experiments
in terms of a  model including an easy plane anisotropy.  This term
in the Hamiltonian accounts for the experimentally observed spin
direction in the crystallographic $a-b$ plane, and it also
contributes to the stability of the quasi-classical ground state.
There are intriguing possibilities for similar measurements on
helimagnets with external field applied in the plane of the spin
rotation.  The theoretical methods applied in this work can not be
readily adapted to this new configuration. It is known that the
static spin order turns into a a soliton
lattice,\cite{maslov,jacobs} but calculations for the ESR
excitations have not been performed yet.

We are indebted to Jessica Thomas for the X-ray
work, G.L. Carr for developing the IR facilities at U12IR, to K.
Holczer for continued discussions and support, and to A. J\'anossy for
consultations on the technical and theoretical aspects of ESR. L.M.
feels honored to receive the Szent-Gy\"orgyi fellowship from the
Hungarian Ministry of Education. B. D. was supported by the Magyary
Zolt\'an postdoctoral program of Magyary Zolt\'an Foundation for
Higher Education (MZFK). Financial support of the Hungarian Research
Founds OTKA TS049881 is acknowledged. The work in Lausanne was
supported by the Swiss NSF, and its NCCR "MaNEP".  Use of the
National Synchrotron Light Source, Brookhaven National Laboratory,
was supported by the U.S. Department of Energy, Office of Science,
Office of Basic Energy Sciences, under Contract No.
DE-AC02-98CH10886.


\end{document}